\documentstyle[12pt,aps]{revtex}
\input epsf
\begin{document}
\title{
\hfill\parbox[t]{2in}{\rm\small\baselineskip 10pt
{~~~~~~JLAB-THY-98-03}\vfill~}
\vskip 0.01cm
\hfill\parbox[t]{2in}{\rm\small\baselineskip 10pt
{revised 28 August 1998}\vfill~}
\vskip 2cm
Duality in Inclusive Semileptonic Heavy Quark Decay}

\vskip 1.0cm

\author{Nathan Isgur}
\address{Jefferson Lab\\
12000 Jefferson Avenue,
Newport News, Virginia 23606}
\maketitle

\vspace{4.0 cm}
\begin{center}  {\bf Abstract}  \end{center}
\vspace{.4 cm}
\begin{abstract}

	I identify a source of $\Lambda_{QCD}/m_Q$ corrections to the assumption of 
quark-hadron duality in the application of heavy quark methods to inclusive
heavy quark semileptonic decays $Q\rightarrow q \ell \bar \nu_{\ell}$.  These corrections 
could substantially affect the accuracy of such methods in practical
applications and in particular
compromise their utility for the extraction of the
Cabibbo-Kobayashi-Maskawa matrix element
$V_{cb}$. 
\bigskip\bigskip\bigskip\bigskip

\end{abstract}
\newpage

  Although the classic application of heavy quark symmetry is in the exclusive 
semileptonic decays of heavy quarks \cite{IWoriginal}, there has also been substantial work on
using heavy quark effective theory (HQET) \cite{HQEToriginal} to systematically improve decay 
predictions for inclusive semileptonic decay rates of heavy hadrons induced
by an underlying $Q\rightarrow q \ell \bar \nu_{\ell}$ 
decay \cite{inclorig,Russianinclusives,otherinclusives}.  In these 
inclusive applications, decays
are treated in an operator product expansion (OPE)  which leads {\it via} HQET
to a $1/m_Q$ expansion in which the leading term is free quark decay,
$1/m_Q$ terms appear to be absent, and terms of order $1/m_Q^2$
take into account such effects as the difference $\bar \Lambda \equiv m_B-m_b$ 
between $m_B$ (the mass of a $\bar B$ meson) and $m_b$ (the mass of a $b$ quark), 
the nonzero kinetic energy ($-\lambda_1 / {2m_b}$) of the heavy quark, small residual
spin-dependent interactions ($2 \lambda_2 / m_b \equiv m_{B^*}-m_B$) of the 
heavy quark, {\it etcetera}.  Although these calculations have become very 
sophisticated \cite{Russianinclusives,otherinclusives}, it is widely 
appreciated \cite{georgicomm,duality} that there remains a basic 
unproved hypothesis in their derivation:  the assumption of quark-hadron duality.
It is the accuracy of this assumption that I want to call into question here.

	 While supposedly valid for any semileptonic decay  $Q\rightarrow q \ell \bar \nu_{\ell}$ of a heavy
quark $Q$, recent applications have centered around the hope that this approach
offers an alternative to the classic exclusive methods
for determining $V_{cb}$, and I will accordingly focus most of my remarks
on the case $b \rightarrow c \ell \bar \nu_{\ell}$ where both quarks
are heavy.  In inclusive $b\rightarrow c \ell \bar \nu_{\ell}$ 
decays, which materialize as $\bar B\rightarrow X_c \ell \bar \nu_{\ell}$, about 65\%
of the $X_c$ spectrum is known to be due to the very narrow ground states $D$ and $D^*$.  The
relatively narrow $s_{\ell}^{\pi_{\ell}}={3 \over 2}^+$ states \cite{IWspec} 
$D_2^*(2460)$ and $D_1(2420)$ account for perhaps
another 5\% of the rate, and it may be assumed that the remaining rate involves decays 
to higher mass resonances (quarkonia and hydrids) and continua \cite{exlusincluscomment}.  The HQET-based 
inclusive calculations predict continuous $X_c$ spectra which are assumed to be
dual to the true hadronic spectrum (see Fig. 1).

  A picture like Fig.~1  might lead one to dismiss the duality approximation
since the inclusive spectrum clearly does not meet the usual requirement
that it be far above the resonance region \cite{masscomment}.  {\it I.e.}, normally the accuracy of 
quark-hadron duality would be determined by a parameter $\Lambda_{QCD}/ E$ where 
the relevent energy scale $E$
is the
mean hadronic excitation energy $\Delta m_{X_c} \equiv \bar m_{X_c}-m_D$.  However, as first explained by Shifman and 
Voloshin \cite{SV,NIonSV}, this is {\it not}
the expansion variable in this case: duality for heavy-to-heavy semileptonic decays sets in {\it 
at threshold} since even as $\delta m \equiv m_b - m_c$ (and therefore $\Delta m_{X_c}$)
approaches zero,
as $m_b \rightarrow \infty$
the heavy recoiling $c$ quark has an energy much greater than $\Lambda_{QCD}$ so that its
final state interactions can be ignored.  In the small velocity (SV) limit, 
it {\it must} therefore hadronize with
unit probability (up to potential $\Lambda_{QCD}/ m_Q$ corrections) as $D$ and $D^*$. 
This ``cannonball" approximation is in fact an essential part of the physical basis of 
the HQET expansion in $1/ m_Q$.  Thus the issue is not whether duality holds in semileptonic
heavy quark decays,  but rather how accurately it holds.

\bigskip

%
%
\begin{center}
~
\epsfxsize=3.0in  \epsfbox{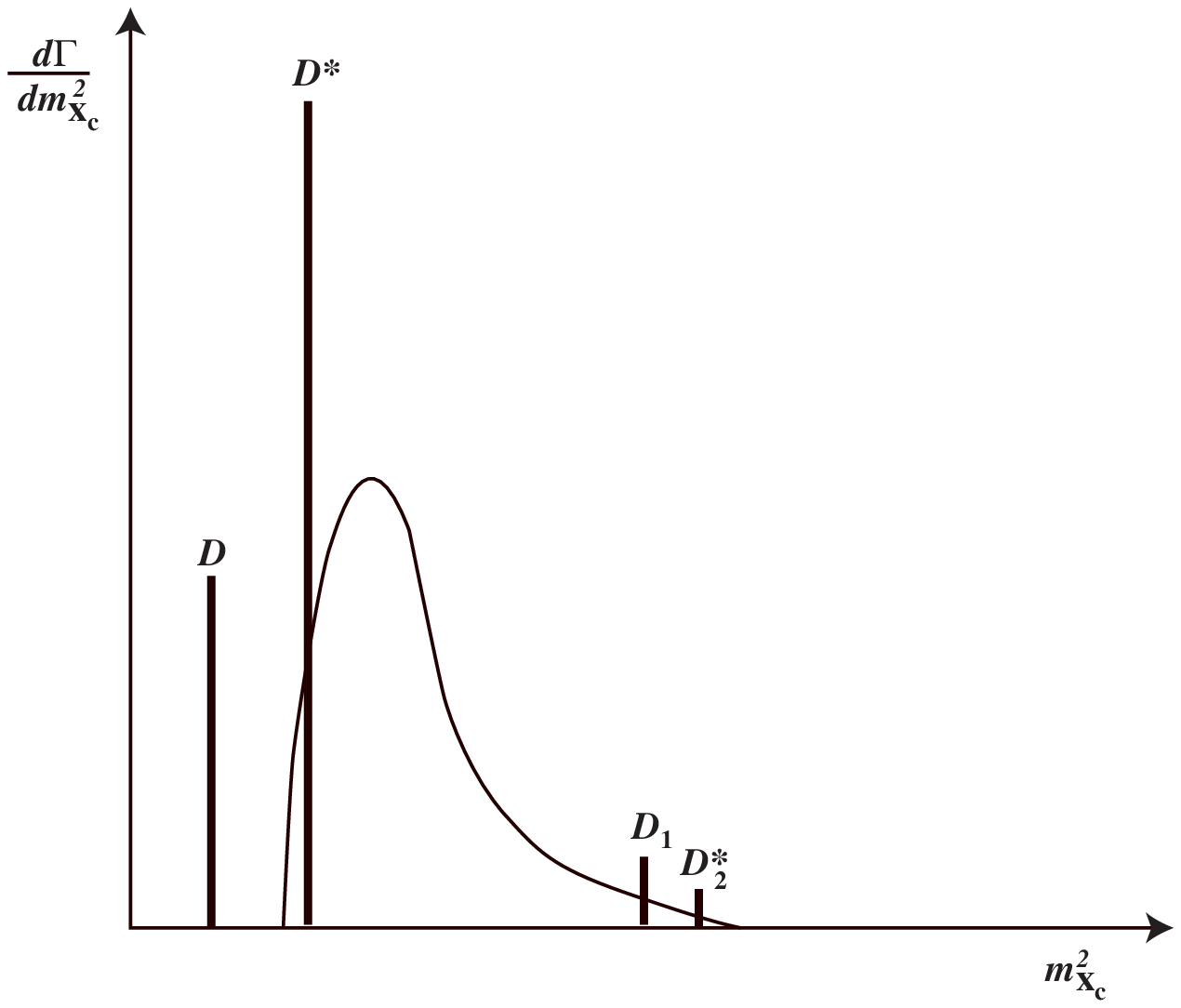}
\vspace*{0.1in}
~
\end{center}

\noindent{ Fig. 1: A sketch for $b \rightarrow c$ semileptonic decay of the continuous inclusive recoil 
spectrum of the OPE calculations (smooth curve) compared to the known hadronic spectrum (shown as
individual resonance lines).}

\bigskip

  Let me begin my discussion of the accuracy of inclusive methods with 
an overview.  Up to {\it caveats} regarding the unknown accuracy of the assumption
of duality, the OPE 
indicates that inclusive calculations should be accurate up to corrections of order $\Lambda^2_{QCD}/
m^2_Q$.  Here I will identify a source of
duality-violation which leads to $\Lambda_{QCD}/m_Q$ corrections.  It is revealed by considerations of  a
Bjorken sum rule \cite{BjSR} which may be viewed as an extension of  Shifman-Voloshin duality to arbitrary 
recoils.  Bjorken's sum rule guarantees that, as $m_b \rightarrow \infty$, duality will be enforced locally in the
semileptonic decay Dalitz plot of rate versus $w-1$ and $E_{\ell}$ (where $w \equiv v \cdot v'$ is the usual heavy 
quark double-velocity
variable and $E_{\ell}$ is the lepton energy).  For regions of the Dalitz plot for which $w-1$ is not
large (and in $b \rightarrow c$ decay nearly the whole Dalitz plot satisfies this condition), the Bjorken sum rule
explicitly relates the loss of total rate from the ``elastic" $s_{\ell}^{\pi_{\ell}}={1 \over 2}^-$ 
channels, as the Isgur-Wise function falls, to the
turn-on of the production of $s_{\ell}^{\pi_{\ell}}={1 \over 2}^+$ and ${3 \over 2}^+$
states \cite{IWonBj}.  

\bigskip\bigskip

%
%
\begin{center}
~
\epsfxsize=3.0in  \epsfbox{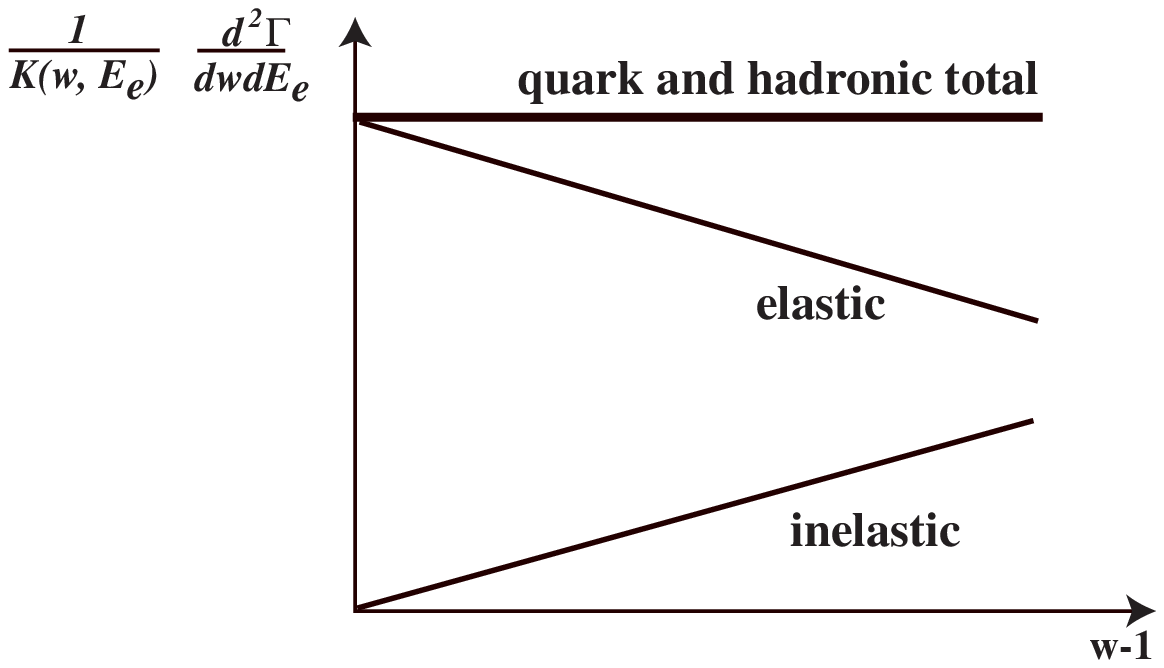}
\vspace*{0.1in}
~
\end{center}

\noindent{ Fig. 2:  The exact compensation by inelastic channels 
of the fall of the elastic rate 
in the linear region as $m_b \rightarrow \infty$.}

\bigskip

   In particular, in this region the Isgur-Wise 
function may be taken to be linear:
\begin{equation}
\xi(w) \simeq 1-\rho^2(w-1) \equiv 1-\left[ {1\over 4}+\rho^2_{dyn} \right] (w-1)~~~.
\end{equation}
In that case, for fixed $r \equiv m_c/m_b$, as $m_b \rightarrow \infty$ inelastic 
$s_{\ell}^{\pi_{\ell}}={1 \over 2}^+$ and ${3 \over 2}^+$ channels 
open up to give a semileptonic rate that would exactly and locally compensate in the Dalitz plot the loss of
rate from the elastic channels due to $\rho^2_{dyn}$.  {\it I.e.}, if
\medskip
\begin{equation}
{d^2\Gamma^{inclusive}_{quark} \over {dwdE_{\ell}}} = K(w,E_{\ell})
\end{equation}
then \cite{BjSR,IWonBj}
\medskip
\begin{equation}
{d^2\Gamma^{inclusive}_{hadron} \over {dwdE_{\ell}}} = K(w,E_{\ell}) \Biggl( {w+1 \over 2} \vert \xi(w) \vert^2
+2(w-1) \left[ \sum_m \vert \tau^{(m)}_{1 \over 2}(1) \vert^2 
+2 \sum_p \vert \tau^{(p)}_{3 \over 2}(1) \vert^2 \right] \Biggr)
\end{equation}
as $m_b \rightarrow \infty$ 
and for $\rho^2 (w-1)_{max} =\rho^2 {(1-r)^2 \over 2r} << 1$.
Since according to (1)
\begin{equation}
\Biggl({w+1 \over 2}\Biggr) \vert \xi(w) \vert^2
\simeq 1- 2 \rho^2_{dyn}  (w-1)~~~,
\end{equation}
we see explicitly the compensation of the dynamical part of the fall of the elastic channels by the 
onset of inelastic states (of all sorts) with 
$s_{\ell}^{\pi_{\ell}}={1 \over 2}^+$ and ${3 \over 2}^+$.  This situation is sketched in Figure 2; if it were
applicable to $b \rightarrow c$ decays, then quark-hadron duality would be exact.

   Having established  conditions for its validity as $m_b \rightarrow \infty$, 
it is easy to see why one should be concerned about 
quark-hadron duality for $b \rightarrow c$ decays.  
For fixed $r$,
$w-1$ lies in the fixed range from $0$ to $(1-r)^2/2r$, and as $m_b \rightarrow \infty$ any given hadronic threshold
collapses to the point $w=1$.  However, for finite $m_b$ there is a gap in $w-1$ in which the rate to the
elastic ${1 \over 2}^-$ channels falls by $\Lambda_{QCD}/m_Q$ terms 
but the potentially compensating excited state channels 
${1 \over 2}^+$ and ${3 \over 2}^+$ are not yet
kinematically allowed.  
More precisely, if $m_{D^{**}}$ is the mass of a generic charmed 
inelastic state, then $t^{**}_m=(m_B-m_{D^{**}})^2$  would be the threshold in $t$ for this state, 
corresponding to a value of $w-1$ 
in the quark-decay Dalitz plot of
\begin{equation}
   { {t_m-t^{**}_m} \over 2m_b m_c} \simeq (1-r) {\Delta \over m_c}
\end{equation}
where $t_m \equiv (m_B-m_D)^2 \simeq (m_b-m_c)^2$ and $\Delta \equiv m_{D^{**}}-m_D$.  Since $\Delta \simeq 500$ MeV 
and $(w-1)_{max} \simeq 0.6$, this region covers more than one third of the Dalitz plot and the compensation
is very substantially delayed:  see Figure 3.  Eqs. (5) and (1) show that this effect is of order
$\Lambda_{QCD}/m_Q$, seemingly at odds with the OPE result.

   Despite this apparent contradiction, there is actually no inconsistency: the OPE
result that the leading corrections to the inclusive rate are of order  $\Lambda^2_{QCD}/m^2_Q$
can still be valid as derived in the limit of large energy release in
the $b \rightarrow c$ transition, while $\Lambda_{QCD}/m_Q$
effects can arise for energy
releases of the order of $\Lambda_{QCD}$ due to 
a finite radius of convergence of the OPE. The main purpose of this paper is indeed
to call attention to this  effect, and to use the quark model
to estimate its importance.

%
%
\begin{center}
~
\epsfxsize=4.0in  \epsfbox{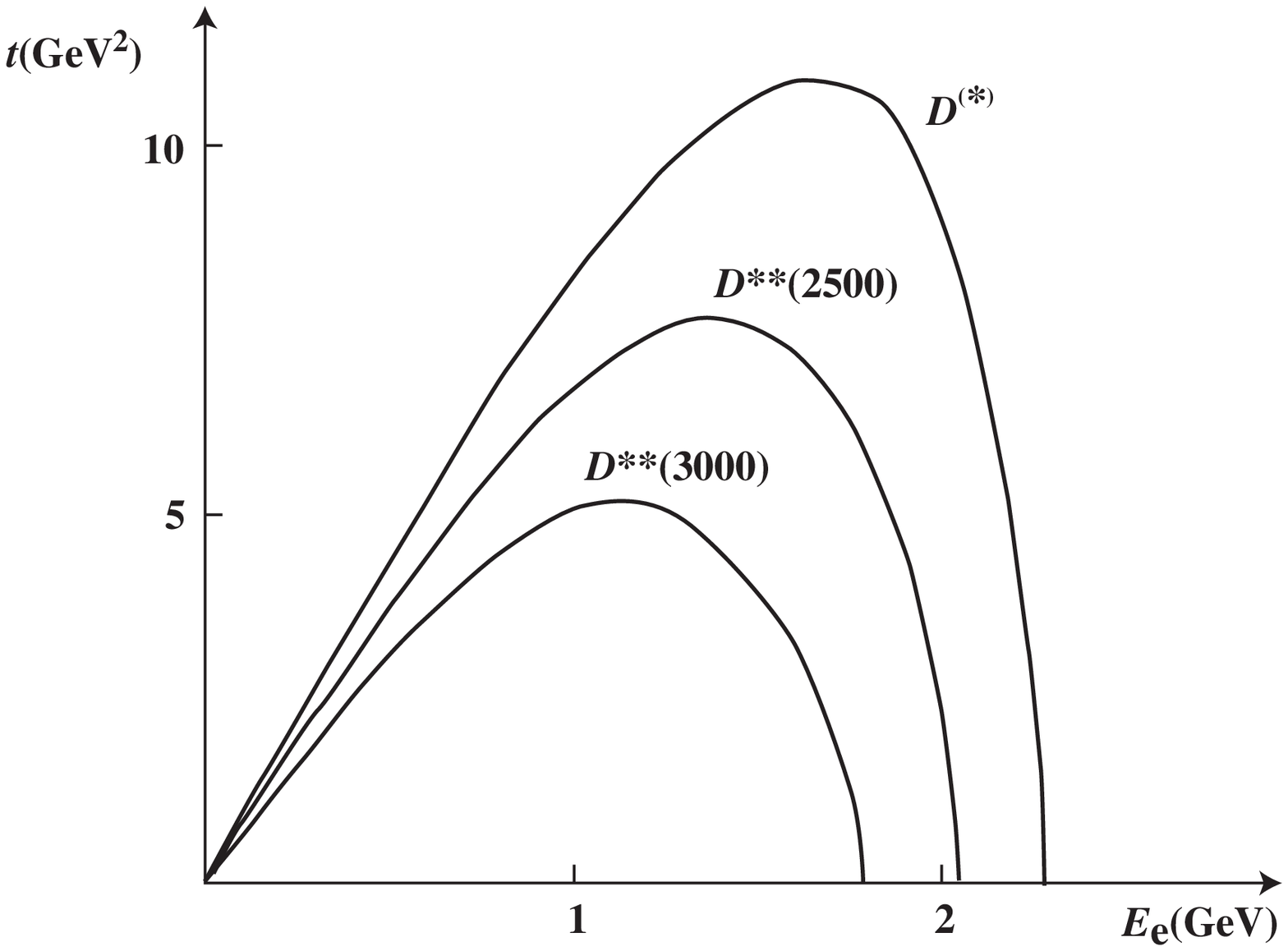}
\vspace*{0.1in}
~
\end{center}

\noindent{ Fig. 3: An overlay of the Dalitz plots for 
$\bar B\rightarrow D^{(*)} e \bar \nu_e$,
$\bar B \rightarrow D^{**}(2500) e \bar \nu_e$,
and $\bar B \rightarrow D^{**}(3000) e \bar \nu_e$. The $D^{(*)}$ mass is taken as
the hyperfine average of the  $D$ and $D^*$ masses; the two
$D^{**}$ masses are chosen for illustrative 
purposes.}

\bigskip

   The basic issues can be most easily exposed by considering \cite{Russianinclusives}
spinless quarks coupled to a scalar field $\phi$ of mass $\mu$, and by studying
the decay $b \rightarrow c \phi$ with weak coupling constant $g$. 
Differential semileptonic decay rates have a more complex spin structure, but otherwise correspond to the case
$\mu=\sqrt{t}$; total semileptonic rates correspond to a weighted average
over kinematically allowed $\mu$ but, as we shall see below, this averaging
does not change the essentials of the problem.
In our simplified case
\begin{equation}
\Gamma (b \rightarrow c \phi) = {g^2p_{cb} \over 8 \pi m_b^2}
\end{equation}
where $p_{fi} \equiv [(m_i-m_f)^2-\mu^2]^{1/2}[(m_i+m_f)^2-\mu^2]^{1/2}/2m_i$
is the momentum of $\phi$ from the two-body decay of mass $m_i$ into masses $m_f$ and
$\mu$. 

    To compare Eq. (6) with a hadronic world, we can use
the quark model framework of Refs. \cite{ISGW,ISGW2} where
\begin{equation}
\Gamma (B \rightarrow D^{(n)} \phi) = {g^2p_{D^{(n)} B} \over 8 \pi m_B^2}
\Bigl({{m_{D^{(n)}}m_B} \over m_cm_b}\Bigr) \vert \xi^{(n)}(\vec v_{D^{(n)} B})\vert^2
\end{equation}
with
\begin{equation}
\xi^{(n)}(\vec v_{D^{(n)} B}) \equiv \int d^3p \phi^*_{D^{(n)}}(\vec p) \phi_B(\vec p -m_d \vec v_{D^{(n)} B})
\end{equation}
and where
\begin{equation}
\vec v_{D^{(n)} B}=\vec p_{D^{(n)} B}/m_{D^{(n)}}
\end{equation}
is the recoil velocity of $D^{(n)}$. These results may be used to estimate
\begin{equation} 
R \equiv {\sum_n \Gamma (B \rightarrow D^{(n)} \phi) \over \Gamma (b \rightarrow c \phi) }~~~.
\end{equation}
Using simple harmonic oscillator wavefunctions in (8) gives
\begin{equation}
\xi_{DB} (\vec v) = \Bigl[ {\beta_D \beta_B \over \bar \beta_{DB}^2}  \Bigr]^{3/2} 
exp \Bigl(-{m_d^2v^2 \over 4 \bar \beta_{DB}^2}\Bigr)
\end{equation}
and
\begin{equation}
\xi^i_{D^{**}B} (\vec v) = {m_dv^i \over \sqrt{2} \beta_B} \Bigl[ {\beta_D \beta_B \over \bar \beta_{DB}^2}  \Bigr]^{5/2} 
exp \Bigl(-{m_d^2v^2 \over 4 \bar \beta_{DB}^2}\Bigr)
\end{equation}
where $\bar \beta_{DB}^2 \equiv {1 \over 2}(\beta_B^2+\beta_D^2)$,
$\xi_{DB}$ is the $B \rightarrow D$ form factor, and
$\xi^i_{D^{**}B}$ is the  form factor for transitions into
the lowest $\ell=1$ excited state with $m_\ell=i$. If we define an order
$\Lambda_{QCD}/m_Q$ expansion parameter 
\begin{equation}
\epsilon \equiv {m_d \over \mu_{-}} = {m_d (m_b-m_c) \over m_b m_c}
\end{equation}
and the ``scaled energy release" 
\begin{equation}
T^* \equiv  {m_b-m_c - \mu \over \Delta}
\end{equation}
where $\Delta \equiv m_{D^{**}}-m_D$, then for small $T^*$ ({\it i.e.}, small velocities) 
\begin{equation}
\vert \xi_{DB} \vert^2 = 1 - \epsilon T^* + O(\epsilon^2)
\end{equation}
and, defining $\vert \xi_{D^{**}B} \vert^2 \equiv \sum_i \vert \xi^i_{D^{**}B} \vert^2$,
\begin{equation}
\vert \xi_{D^{**}B} \vert^2 = \epsilon (T^*-1) + O(\epsilon^2)~~~.
\end{equation}
Since in this limit
\begin{equation}
{p_{D^{(n)}B} \over p_{cb}} = \Bigl[ {m_{D^{(n)}} m_b \over m_c m_B} \Bigr]^{1/2} 
\Bigl({T^*-1 \over T^*} \Bigr)^{1/2}~~,
\end{equation}
we can obtain a model \cite{Russianinclusives} for $R$ by truncating the sum over $n$ after the first $D^{**}$:
\begin{equation} 
R_1^{D^{**}} \equiv {\Gamma (B \rightarrow D \phi)+\Gamma (B \rightarrow D^{**} \phi) 
\over \Gamma (b \rightarrow c \phi) }
\end{equation}
\begin{equation} 
~~~~~~~=  [1 +{3\over 2} \epsilon - \epsilon T^*] \theta (T^*)
+ \epsilon {(T^*-1)^{3/2} \over {T^*}^{1/2}} \theta (T^*-1)~~~,
\end{equation}
wherein we have made the weak binding approximation $\beta << m_d$
and shown explicitly the two thresholds at $T^*=0$ and $T^*=1$.
As $T^* \rightarrow \infty$, the complete expression for $R$ will display a tower of 
states (each with their appropriate threshold factors)  being produced with a strength
given by a
power series in $T^*$  with coefficients which are 
power series in generalized $\epsilon$-coefficients of order $\Lambda_{QCD}/m_Q$.
For duality to be valid, each term $T^{*n}$ for $n>0$ must
have zero coefficient.  The OPE further states that the constant
term $T^{*0}$ is of the form  $1+O(\Lambda_{QCD}^2/m_Q^2)$.

	  While extreme, this  truncation of the complete expression for 
$R$ has the properties that:

1.	At $T^* \rightarrow \infty$, it is of the form 
$1+O(\epsilon^2)+O(\epsilon/T^*)$ as required by the OPE.

2.	There are no other terms of order $1$,
$\epsilon$, or $\epsilon T^*$ 
possible beyond those shown:  a more accurate treatment
of $\Gamma (B \rightarrow D \phi)$ could only generate
$\epsilon^2$, $\epsilon^2 T^*$, $\epsilon^2 T^{*2}$,~...
terms;  a more accurate treatment of $\Gamma (B \rightarrow D^{**} \phi)$ could only generate
$\epsilon^2 T^*$, $\epsilon^2 T^{*2}$,~... terms; and all 
higher states first make a contribution at order $\epsilon^2 T^{*2}$ or higher.  Conversely,
we note that if, for example, $\epsilon^2 T^{*2}$ terms are 
retained, they must all cancel exactly or the requirements of the OPE would be violated
as $T^* \rightarrow \infty$.

3.	As $\Delta m \equiv m_b-m_c \rightarrow 0$, 
$R_1^{D^{**}} \rightarrow 1+O({ \Lambda_{QCD}\Delta m  \over m_b^2})$ as required \cite{SVcomment}.

4.	Near $T^*_{max} \equiv m_b-m_c$, $\epsilon T^*_{max} $ is in general large.  
This observation corresponds in the usual language of heavy quark symmetry to the statement that 
the natural scale of the slope $\rho^2$
of the Isgur-Wise function is of order unity.  It is
also consistent with the experimental observation that 
$\vert \xi_{DB}  \vert^2$ has dropped to less than half its value between $w-1=0$
and its maximum value.  Given this, the extension of Eq. (19) to higher orders in 
$T^*$ will require
a ``conspiracy" of the entire spectrum of possible hadronic final states.  Nevertheless, we can use
Eq. (19) across the full range of $T^*$ as an indicator of the 
$\Lambda_{QCD}/m_Q$ effects arising from the order 1 and order $T^*$ terms in the
expansion of $R$; this corresponds to a ``best case" assumption that 
duality is locally perfect for the terms $T^{*n}$
with $n>1$.

	   While this simple example clearly demonstrates the existence of the claimed duality-violating 
$\Lambda_{QCD}/m_Q$ effects  for finite $T^*$, it 
remains to discuss their quantitative importance.  Before doing so, I will introduce a number of simple variants of this 
prototypical model.  The first corrects an idiosyncracy of the simple harmonic oscillator model: in it the 
first $D^{**}$ state saturates the Bjorken sum rule, {\it i.e.}, completely compensates the 
$-\epsilon T^*$ elastic term.  In contrast, in the
ISGW model where harmonic oscillator wavefunctions are simply used as a variational basis, 
$\beta_{D^{**}} \neq \beta_D$ and this 
saturation does not occur.  This is typical of the general case where (in the narrow resonance approximation) 
the total resonant 
$P$-wave term is of the form
\begin{equation} 
{\epsilon \over {T^*}^{1/2}} \sum_n f_n (T^*-t^*_n)^{3/2}  \theta (T^*-t^*_n)
\end{equation}
with $\sum_n f_n=1$ and $t^*_n$ being the threshold for channel $n$.
As $T^* \rightarrow \infty$, these contributions automatically  cancel the 
$-\epsilon T^*$ term from the elastic form factor, and constrain the $O(\epsilon)$
correction:
\begin{equation} 
R_{1+2+...}^{D^{**}} =  [1 +{3\over 2} \epsilon \bar t^* - \epsilon T^*] \theta (T^*)
+ {\epsilon \over {T^*}^{1/2}} \sum_n f_n (T^*-t^*_n)^{3/2}  \theta (T^*-t^*_n)~~~,
\end{equation}
where
\begin{equation} 
\bar t^* = \sum_n f_n t^*_n
\end{equation}
is the weighted average threshold position.  Note that since some $T^*_n$ exceed
$T^*_{max}$, $R_{1+2+...}^{D^{**}}$ cannot  heal to unity in the physical
decay region.

	As described above, both 
$R_{1}^{D^{**}}$ and 
$R_{1+2+...}^{D^{**}}$ are ``best case" truncations which assume exact cancellations of 
$\epsilon^2 T^*$, $\epsilon^2 T^{*2}$,~... terms.  While sufficient
for the purposes of this study, I note that it is straightforward to recursively ``construct duality" to the
required order in $\epsilon$ to any finite order in $T^*$.  For example, for a simple harmonic oscillator
spectrum one can easily construct
\begin{eqnarray} 
R^{ho} = {exp(-\epsilon T^*) \over {T^{*1/2}}} 
&&\Bigl(
[1 +{3\over 2} \epsilon]  {T^*}^{1/2} \theta (T^*)\nonumber \\
&&+\epsilon[1 +{5\over 2} \epsilon +{35 \over 16}\epsilon^2 +{35 \over 32}\epsilon^3 +{385 \over 1024}\epsilon^4 + ...]
  (T^*-1)^{3/2} \theta (T^*-1)\nonumber \\
&&+{1 \over 2!}\epsilon^2[1 +{7\over 2} \epsilon +{21 \over 4}\epsilon^2 +{77 \over 16}\epsilon^3 + ...] 
  (T^*-2)^{5/2} \theta (T^*-2)\nonumber \\
&&+{1 \over 3!}\epsilon^3[1 +{9\over 2} \epsilon +{297 \over 32}\epsilon^2 +...]  (T^*-3)^{7/2} \theta (T^*-3)\nonumber \\
&&+{1 \over 4!}\epsilon^4[1 +{11\over 2} \epsilon + ...]  (T^*-4)^{9/2} \theta (T^*-4)\nonumber \\
&&+{1 \over 5!}\epsilon^5[1+ ...] (T^*-5)^{11/2} \theta (T^*-5)\nonumber + ...
\Bigr)~~~,
\end{eqnarray}
where the ellipses denote terms of order $\epsilon^6 {T^*}^n$ with $1 \leq n \leq 5$ and all terms of
order $\epsilon^m {T^*}^m$ and higher with $m > 5$.  This truncated expansion is 
accurate even at $T^*=5$ up to corrections of order $\Lambda_{QCD}^2/m_Q^2$; as we will see 
below, higher values of $T^*$ are probably not physically 
relevant.

	   The models just introduced are all based on the duality of $b \rightarrow c \phi$
to a tower of $c \bar d$ resonances.  While the thresholds
associated with such towers are a source of duality-violating 
$\Lambda_{QCD}/m_Q$ corrections which must be a cause for concern
in comparing inclusive calculations with experiment, I am even more concerned about 
processes which could give a nonperturbative high-mass tail to the recoil mass distribution. 
The convergence to unity of $R$  in the former case
would be controlled by an expansion in $1/T^* \sim \Delta/(m_b-m_c-\mu)$, {\it i.e.}, 
$T$ must be large compared to the single resonance scale $\Delta$.  Since $\Delta/T_{max} \sim 1/6$, 
there are reasons to be cautious, but perhaps not alarmed.  However,
I believe that there is another ``harder" effective scale in low energy hadron structure which  I identify
with the constituent quark size $r_q$. While, like $\Delta$, $1/r_q$ must be ``of order $\Lambda_{QCD}$", 
empirically it is much larger:  the constituent quark model makes sense 
only if the quarks are small relative to hadronic radii.
Quantitative estimates from spectroscopy \cite{quarkradius} indicate that this scale is indeed 
$1/r_q \sim 2$ GeV.  This is in turn potentially very 
dangerous for duality when $T_{max}$ is only of order 3 GeV as in $b \rightarrow c$ decays.  

      Let me mention two concrete examples of how this scale could 
lead to a high-mass tail to the recoil mass distribution which would delay the compensation
required for duality. If the glue around a constituent quark is indeed very compact, then
high recoil momenta of the $c$ quark are required if it is to be left behind, {\it i.e.}, before the 
current $c$ quark can carry away all of the energy of the underlying $b \rightarrow c \phi$ transition.
This effect would lead to a 
convergence problem for the resonance
contributions to $R$ associated with the excitation of hybrid mesons.
The second example is probably more dangerous:  such a scale could
lead to the nonperturbative production of high-mass
nonresonant states \cite{nonresonant}, {\it e.g.}, an $X_cY$ mass spectrum for 
$B \rightarrow X_cY\phi$  extending from $X_cY$ threshold up to masses of order 
$m_{X_c}+m_Y+1/r_q \sim 5$ GeV $\sim m_B$.  We should 
therefore be concerned that a substantial fraction of nonresonant production is unavailable
to participate in the compensation
required for duality to be realized in $b \rightarrow c$ decays.

	   For a crude estimate of the effects of a high-mass hybrid contribution, I take a simple two-component resonance model consisting
of ``normal" $c \bar d$ resonances with $\bar t^*_{c \bar d}$  and $c \bar d$ hybrids with 
$\bar t^*_{hybrid}$ substantially larger.  If we assume that the
latter are responsible for a fraction $\kappa$ of $\rho^2_{dyn}$, then we would have
\begin{eqnarray} 
R^{hybrid} &=&  [1 +{3\over 2} \epsilon \bar t^* - \epsilon T^*] \theta (T^*) \nonumber \\
&&+ (1-\kappa)\epsilon {(T^*-\bar t^*_{c \bar d})^{3/2} \over {T^*}^{1/2}} \theta (T^*-\bar t^*_{c \bar d}) \nonumber \\
&&+ \kappa\epsilon {(T^*-\bar t^*_{hybrid})^{3/2} \over {T^*}^{1/2}} \theta (T^*-\bar t^*_{hybrid})~~~,
\end{eqnarray}
with $\bar t^*=(1-\kappa) \bar t^*_{c \bar d}+ \kappa \bar t^*_{hybrid}$.  
Since both experimentally and theoretically the hybrid mass spectrum begins about $1.5$ GeV above the ground state,
this provides a minimum value for $\bar t^*_{hybrid}$. Since their ``hard" production
mechanism will raise their mean $T^*$ above this minimum,
the effects of their postponed onset could be serious.

	As already implied, I believe that the postponed nonresonant contributions are an even more serious
cause for concern.  While the
model (23) for the hybrids could also be used as a template for a crude model 
for nonresonant states (their contribution will 
in first order be controlled by a parameter $\bar t^*_{nr}$ which would replace 
$\bar t^*_{hybrid}$ in Eq. (23)), a more appropriate model would be
\begin{eqnarray} 
R^{nr} &=&  [1 +{3\over 2} \epsilon \bar t^* - \epsilon T^*] \theta (T^*) \nonumber \\
&&+ (1-\lambda)\epsilon {(T^*-\bar t^*_{c \bar d})^{3/2} \over {T^*}^{1/2}} \theta (T^*-\bar t^*_{c \bar d}) \nonumber \\
&&+ \lambda\epsilon \int_{T^*_{min}}^{T^*}dt^* \rho (t^*){(T^*-\bar t^*)^{3/2} \over {T^*}^{1/2}} ~~~,
\end{eqnarray}
where $\lambda$ is the fraction of $\rho^2_{dyn}$ due to nonresonant states 
and $\rho(t^*)$ is the appropriate normalized spectral function ($\int_{T^*_{min}}^{\infty}dt^* \rho (t^*)=1$)
which begins at $T^*_{min}$
but drops off very slowly with a scale determined by $1/r_q$.  In this situation,
$\bar t^*=(1-\lambda) \bar t^*_{c \bar d}+ \lambda \int_{T^*_{min}}^{\infty}dt^* \rho (t^*)t^*$.

~
\bigskip\bigskip
~
%
%
\begin{center}
~
\epsfxsize=6.0in  \epsfbox{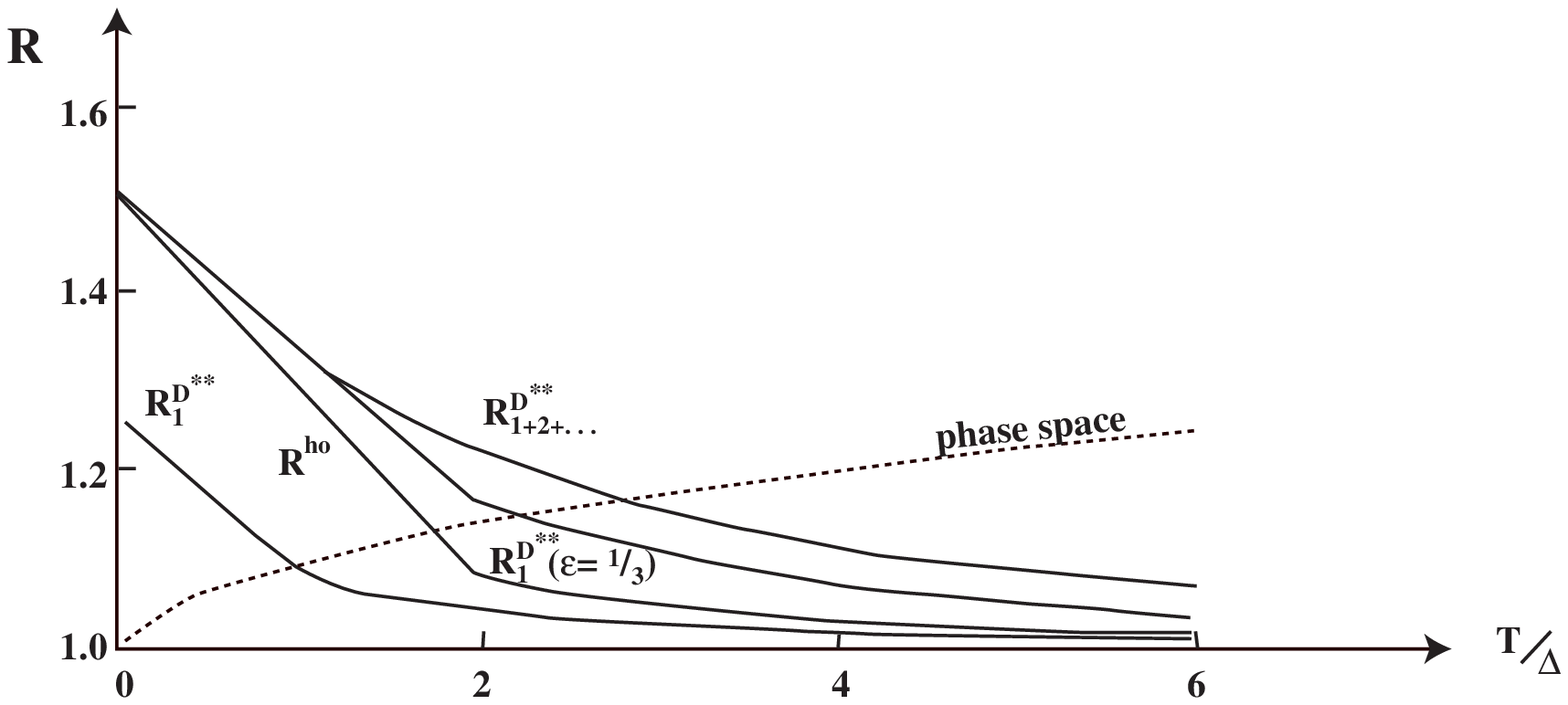}
\vspace*{0.1in}
~
\end{center}

\bigskip

\noindent{ Fig. 4:  Four resonance models of the approach to duality:
(a) $R^{D^{**}}_1$ (with the baseline value $\epsilon=1/6$),
(b) $R^{D^{**}}_1$ (with $\epsilon=1/3$ corresponding
to $T^*_{max}=3$),
(c) $R^{D^{**}}_{1+2+ ...}$ (with $t^*_n=n$ and $f_n=({1 \over 2})^n$ so that $\bar t^*=2$), and
(d) $R^{ho}$.}

\bigskip\bigskip

	I now turn to rough quantitative estimates of  duality-violating 
effects using the models I have presented.  Since
experimentally $\rho^2 \sim 1$, $(w-1)_{max} \sim 1/2$ and $(m_b-m_c)/\Delta \sim 6$, 
I will take $\epsilon=1/6$ and $T^*_{max}=6$ as a realistic baseline for all of the following examples.
However, in keeping with the main message of this paper that
inclusive results must
be interpreted cautiously until duality-violating effects are better understood,
I will choose pessimistic
values for other parameters of the models and for variations around these baseline values.

	Figure 4 shows four different examples of resonance compensation:
$R^{D^{**}}_1$ (with both the baseline value $\epsilon=1/6$ and for $\epsilon=1/3$ corresponding
to $T^*_{max}=3$, {\it i.e.}, to using $\Delta_{eff}=2 \Delta$ for the mean location of the P-wave
strength), $R^{D^{**}}_{1+2+ ...}$ (with $t^*_n=n$ and $f_n=({1 \over 2})^n$ so that $\bar t^*=2$),
and $R^{ho}$.
These examples show that duality-violation of the order of 10\% could easily arise in $b \rightarrow c$  
decays from the delayed onset
of resonances with a scale $\Delta \simeq 500$ MeV.

\bigskip

%
%
\begin{center}
~
\epsfxsize=6.0in  \epsfbox{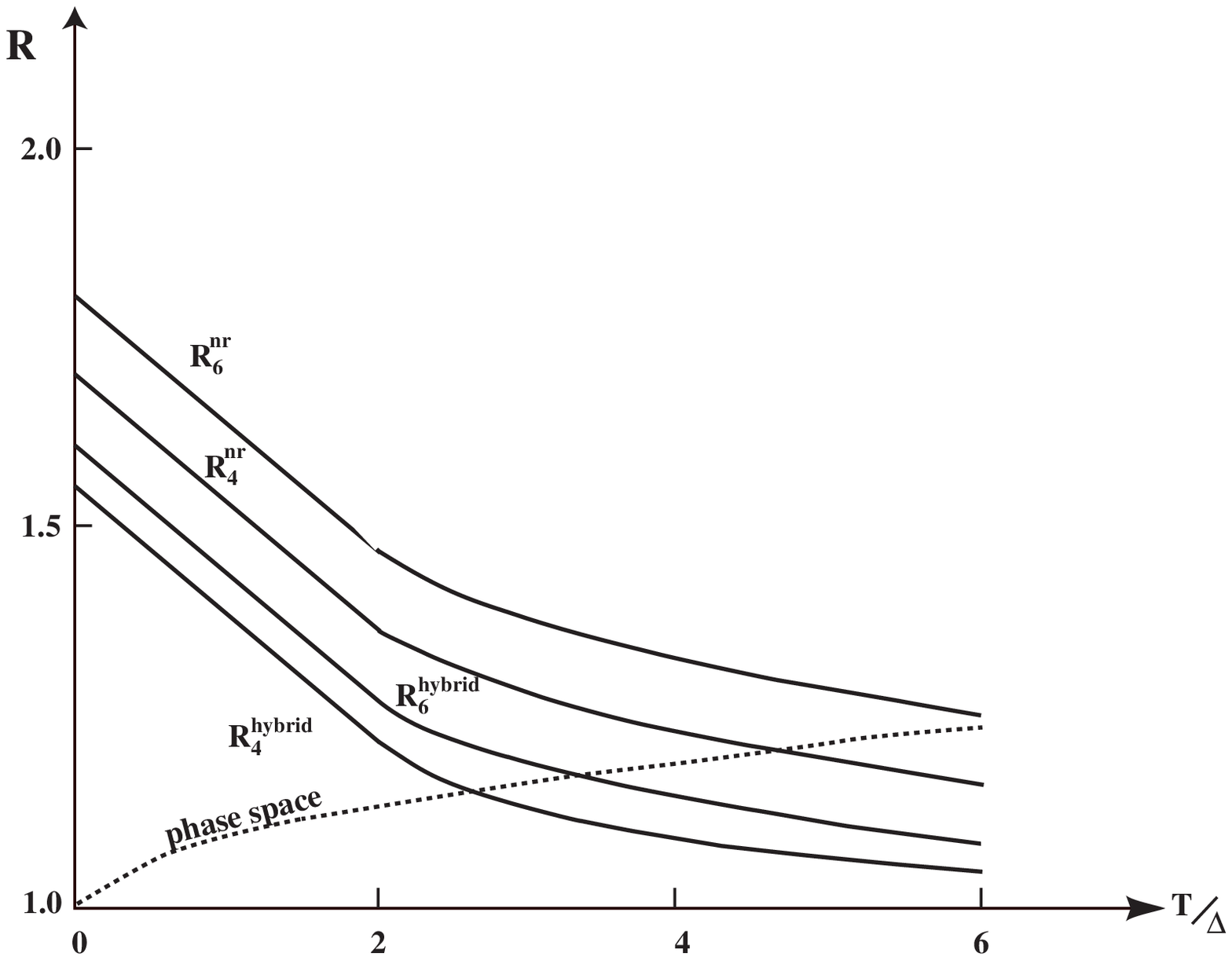}
\vspace*{0.1in}
~
\end{center}

\bigskip

\noindent{ Fig. 5: Four examples of the effects of a nonperturbative high mass tail on the recoil mass spectrum:
(a) $R^{hybrid}$ (with $\kappa=1/10$, $\bar t^*_{c \bar d}=2$,   and $\bar t^*_{hybrid}=4$),
(b) As in (a), but with $\bar t^*_{hybrid}=6$,
(c) $R^{nr}$ (with $\lambda=1/5$, $\bar t^*_{c \bar d}=2$,  $T^*_{min}=2$, and $s=4$), and
(d) As in (c) but with $s=6$.}

\bigskip

	Figure 5 shows the potentially even more dangerous effects of there being a second scale in this problem
much larger than $\Delta$.   Here I once again give four examples:
$R^{hybrid}$ (with $\kappa=1/10$, $\bar t^*_{c \bar d}=2$,   and $\bar t^*_{hybrid}=4$ and 6) and
$R^{nr}$ (with $\lambda=1/5$, $\bar t^*_{c \bar d}=2$, and 
\begin{equation}
\rho(t^*)={1\over s}exp({T^*_{min}-t^* \over s})\theta(t^*-T^*_{min})
\end{equation}
with $T^*_{min}=2$ and $s=4$ and 6). 
Note that the choices made  for 
$\bar t^*_{hybrid}$ and $s$ are based on the hypothetical scale $1/r_q \sim 2$ GeV.  
The  choices $\kappa=1/10$ and
$\lambda=1/5$ are based on estimates \cite{exlusincluscomment} of the strengths of these 
contributions, but values of this order are certainly reasonable ({\it e.g.}, $\lambda$
is a $1/N_c$ effect).  These examples show that it
is difficult to avoid the conclusion that nonperturbative high mass contributions to 
$\rho^2$ could lead to substantial duality-violating $\Lambda_{QCD}/m_Q$
effects in $b \rightarrow c$ decays.

   Although our main focus has been on heavy-to-heavy transitions, the  
physics issues raised here (if not their explicit forms) are also
relevant for  $Q\rightarrow q \ell \bar \nu_{\ell}$ transitions.  
Before concluding, let me therefore point out a simple application of the OPE to inclusive
heavy-to-light transitions where it seems certain to me that they will fail:  Cabibbo-forbidden charm decays. (Even though such decays
might be an unimportant application of the inclusive calculations in practice, they provide a valid theoretical testing
ground for their accuracy.)  In particular, consider the $c\rightarrow d \bar \ell \nu_{\ell}$ 
decays of the $D^0$ and $D^+$.  They will be dominated by 
the channels $D^0 \rightarrow \pi^- \bar \ell \nu_{\ell}$  and $\rho^- \bar \ell \nu_{\ell}$ and by 
$D^+ \rightarrow \pi^0 \bar \ell \nu_{\ell}$, $\eta \bar \ell \nu_{\ell}$, 
$\eta ' \bar \ell \nu_{\ell}$, $\rho^0 \bar \ell \nu_{\ell}$, and $\omega \bar \ell \nu_{\ell}$.  
Since the OPE corrections in the $D^0$ and $D^+$ are {\it identical}, their Cabibbo-forbidden 
semileptonic partial widths and spectral distributions are predicted to be identical.  However, simple isospin symmetry implies that 
$\Gamma (D^+ \rightarrow \pi^0 \bar \ell \nu_{\ell})= {1 \over 2} \Gamma(D^0 \rightarrow \pi^- \bar \ell \nu_{\ell})$, 
so the 
inclusive Cabibbo-forbidden rates can only be equal if 
$\Gamma (D^+ \rightarrow \eta \bar \ell \nu_{\ell})+\Gamma(D^+ \rightarrow \eta ' \bar \ell \nu_{\ell})
= \Gamma(D^+ \rightarrow \pi^0 \bar \ell \nu_{\ell})$.  In many models this latter relation 
would be true if $m_{\eta}=m_{\eta '}=m_{\pi}$, since it is rather
natural for the squares of matrix elements to satisfy its analogue.  However, with 
real phase space factors, this relation is typically
badly broken. Since  Cabibbo-forbidden
decays, like their Cabibbo-allowed counterparts, will receive little P-wave compensation,
I expect this prediction to fail.

   Finally, I note that the duality-violating effects I have 
highlighted here will have an effect on the long-standing 
$\bar B$ semileptonic branching ratio puzzle \cite{BRproblem}. 
Since the hadronic
mass distribution in $b \rightarrow c \bar u d$ is weighted toward higher
masses than the leptonic mass distribution in 
$b \rightarrow c \ell \bar \nu_{\ell}$, the ratio
of these two rates will be changed.
    
   In summary, I have shown here that hadronic thresholds lead to $\Lambda_{QCD}/m_Q$
violations of duality in $b \rightarrow c$ decays which  do not explicitly appear
in the operator product expansion. Since such violations cannot appear as the
$b \rightarrow c$ energy release $T \rightarrow \infty$,
there are ``conspiracies"  ({\it i.e.}, sum rules) which relate
hadronic thresholds and transition form factors. As emphasized by Bigi, Uraltsev, Shifman,
Vainshtein, and others \cite{Russianinclusives,otherinclusives,duality}, these relations tend to compensate
the otherwise extremely large $\Lambda_{QCD}/m_Q$ effects even at small $T$. In this paper
I have displayed several models of such hadronic compensation mechanisms which
indicate that these duality-violating $\Lambda_{QCD}/m_Q$ effects could nevertheless be very substantial.
While the examples I have selected are perhaps pessimistic,
they indicate that these effects
must be better understood before inclusive methods can be 
applied with confidence to heavy quark semileptonic decays.

\bigskip\bigskip

\noindent{\bf ACKNOWLEDGEMENTS}

    I am very grateful to  number of colleagues who examined and commented upon
a draft version of this paper, including Adam Falk, Mark Wise, Zoltan Ligeti, Iain Stewart,
Matthias Neubert, and Richard Lebed.

    I am particularly indebted to Ikaros Bigi, Misha Shifman, Nikolai Uraltsev, and Arkady Vainshtein
who replied with a detailed and very pedagogical explanation of
their understanding of the effects discussed in this paper. I
certainly learned more from this exchange than they did, and substantial errors in the draft
version were corrected as a result.

\vfill\eject


\begin{references}
\bibitem{masscomment} As argued in Refs. \cite{ISGW,ISGW2} by considering
the free recoil of the final state quark $c$ against a static spectator $\bar d$ quark, $\Delta m_{X_c}$
should be roughly $[(m_c+m_d)^2+m_dm_b^{-1}(m_b-m_c)^2]^{1\over 2}-(m_c+m_d) \simeq  0.2$ GeV, 
consistent with the data and the inclusive models shown in 
Figure 1, and certainly not leading to a small expansion parameter $\Lambda_{QCD}/
\Delta m_{X_c}$ for $b \rightarrow c$ decays.
\end{references}
\end{document}